\documentclass[12pt,preprint]{aastex}

\shorttitle{Hydrodynamic Mixing and SN Progenitors}
\shortauthors{Young et al.}

\newcommand{\sol}{$M_\odot$}
\def \nuc#1#2{\relax\ifmmode{}^{#1}{\protect\text{#2}}\else${}^{#1}$#2\fi}

\begin{document}

\title{The Impact of Hydrodynamic Mixing on Supernova Progenitors}

\author{Patrick A. Young\altaffilmark{1,2}, Casey Meakin\altaffilmark{2}, David Arnett\altaffilmark{2}, \& Chris L. Fryer\altaffilmark{1,3}}
\altaffiltext{1}{Theoretical Astrophysics, Los Alamos National Laboratories, Los Alamos, NM 87545}
\altaffiltext{2}{Steward Observatory, University of Arizona, 
 Tucson AZ 85721}
\altaffiltext{3}{Physics Dept., University of Arizona, 
 Tucson AZ 85721}
\email{payoung@lanl.gov, cmeakin@as.arizona.edu, darnett@as.arizona.edu, fryer@lanl.gov}

\begin{abstract}
Recent multidimensional hydrodynamic simulations have demonstrated the
importance of hydrodynamic motions in the convective boundary and
radiative regions of stars to transport of energy, momentum, and
composition. The impact of these processes increases with stellar
mass. Stellar models which approximate this physics have been tested
on several classes of observational problems. In this paper we examine
the implications of the improved treatment on supernova
progenitors. The improved models predict substantially different
interior structures. We present pre-supernova conditions and simple
explosion calculations from stellar models with and without the
improved mixing treatment at 23 \sol. The results differ
substantially.
\end{abstract}


\section{INTRODUCTION}

The predicted nucleosynthetic yields of a stellar population, as well
as initial mass functions (IMFs), are dependent on stellar
models. Should the underlying assumptions in a stellar evolution code
change, we would expect the stellar population we infer from an
observed abundance pattern or luminosity function to change as
well. Conversely, predictions of the chemical evolution of galaxies
will also vary.

Multidimensional hydrodynamic simulations of stellar interiors display
important physical processes that are missing from formulations of
stellar evolution. Bulk fluid motions in convective regions create an
unstable boundary layer and excite internal waves which give rise to
mixing and transport of energy in adjacent stably startified layers
\citep{press81,ya05}. As convective plumes enter a region of the star
which is stable against oscillations (the
Br\"{u}nt-V\"{a}is\"{a}l\"{a} frequency $N^2 > 0$)the overlying
material is not engulfed by the plume and accelerated as in the
convective region. Instead, the material undergoes a Lagrangian
displacement and oscillates around its point of origin. The plume
itself deposits its radial energy of motion into this displaced layer
and spreads beneath it. The spreading of the plume and the horizontal
propagation of the waves inject shear into the boundary layer. Plume
impact is a stochastic process which injects energy into internal
waves with a broad superposition of modes. (A detailed and excellent
discussion of wave excitation can be found in
\citet{fva98,ns01,sn01}. As the fraction of pressure contributed by
radiation increases, the perturbations are subjected to less restoring
force. Thus the importance these processes increases with increasing
stellar mass \citep{ykra03,ya05}. For supernova progenitors the effect
can be dramatic.

In this paper we present simple explosion calculations for an initial
mass of 23 \sol from models with and without the contributions from
these processes, which we will call ``hydro mixing'' as a shorthand,
implemented as in \citet{ya05}. In Section 2.1 we present interior
conditions for each of the models. Section 2.2 describes hydro
simulations of the O and C burning shells and their
implications. Section 3 compares the explosions and discusses
additional issues to be considered in a realistic supernova
calculation.

\section{PRE-SUPERNOVA CONDITIONS}

\subsection{Initial Models}

We examine an initial mass of 23 \sol with and without hydro
mixing. This mass is interesting for nucleosynthesis, being relatively
numerous in the IMF while still ejecting a large amount of processed
material per star into the interstellar medium (ISM)
\citep{arnett96}. This is also the mass of the primary member of the
eclipsing binary EM Car. Apsidal motion of the binary
gives us a measurement of size of the convective core on the main
sequence \citep{ymal01,ya05}. The hydro mixing model predicts the
convective core size well, giving us
one constraint on core size. We have performed 2 and 3-D simulations of the
main sequence convective core and the oxygen and carbon burning shells
for this mass (Meakin \& Arnett 2005a,b, in prep.).

All evolution calculations were performed with the TYCHO code
\citep{ya05}. Both models use the solar composition of
\citet{gs98}. Though this abundance pattern looks to be superseded by the
(quite different) values of recent determinations \citep[][and
references therein]{as05}, we choose to use it for comparison
with earlier results. We are most interested in the comparison between
the models, since a quantitatively accurate explosion requires a
multidimensional calculation.

\subsection{Oxygen Shell Burning Simulations}

In this section we show similarities between multi-D hydro
calculations and the 1D TYCHO models and discuss additional features
of the stellar structure apparent only in the hydro simulations. The
simulations help test the assumption of our 1D formulation as well as
identifying new processes. We find the 1D treatment to be
robust. These calculations extend the work of \citet{ba98,sa00} to
higher resolution, improved inital conditions, and 3D.

The simulations used in this comparison were produced with PROMPI, a
version of the PROMETHEUS PPM hydro code parallelized using domain
decomposition for MPI platforms. PROMPI includes the OPAL opacities
and the TYCHO equation of state and nuclear reaction subroutines with
a 25 element reaction network that reproduces the energy generation of
the full 177 element network to $<1\%$. TYCHO models of 23 \sol with
and without hydro mixing were used as initial conditions. A detailed
description of both main sequence and O shell burning calculations
will appear in separate papers (Meakin \& Arnett 2005a,b, in
prep.). Here we summarize the results from O shell models and
concentrate on their consequences for pre-supernova
models. Simulations were run for 2D wedges encompassing the O shell
and stable regions on either side for both types of initial models
(ob.2d.c and ob.2d.m). A 3D wedge (ob.3d.B) was also run for the
standard initial model. Table~\ref{tab1} summarizes this subset of
models from a larger study, with inner and outer radius, angular
extent of wedge, number of zones in each dimension, and length of the
simulation.

\begin{figure}
\figurenum{1}
\includegraphics[scale=0.7]{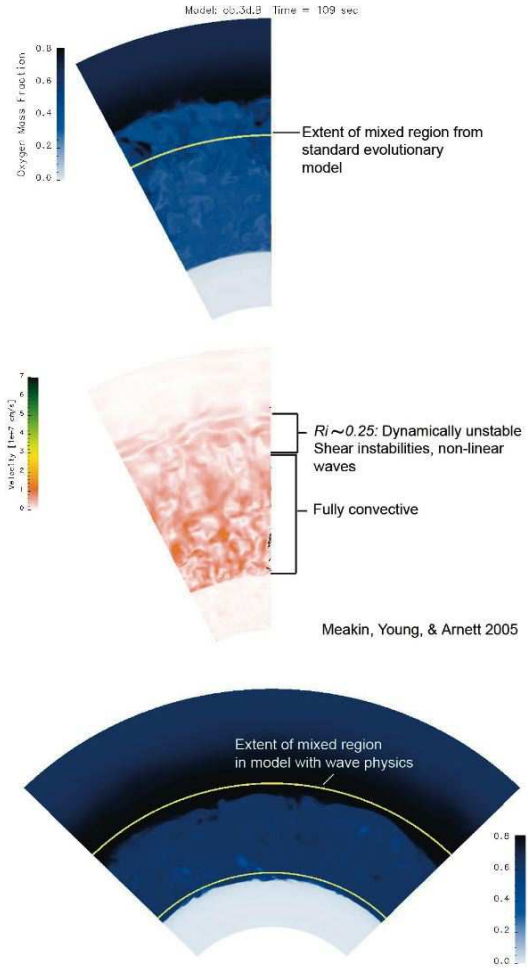}
\caption{O mass fraction (top) and velocity (middle) for the 3D wedge
with standard initial model (ob.3d.B) and O mass fraction for a 2D
wedge with hydro mixing initial model (ob.2d.m) (bottom). The yellow
line on the top panel denotes the extent of the mixed region in the 1D
standard model. When convection develops, the mixed region extends
itself well past the original boundary. The velocity in the extended
mixing region has a different character from that of the convective
region below. In the 2D wedge with hydro mixing initial model
(bottom), the mixed region stabilizes near the initial model boundary
(yellow). The mixed region extends past the lower boundary because of
a resolution effect.\label{fig1} }
\end{figure}

Figure~\ref{fig1} shows O mass fraction (top) and velocity (middle)
for the 3D wedge (ob.3d.B). The yellow line on the top panel denotes
the extent of the mixed region in the standard model. As soon as
convection develops, the mixed region extends itself well past the
original boundary and stabilizes at the new size.  The same behavior
of is observed in 2D (ob.2d.c) and 3D. The fluid velocities in the extended
mixing region have a different character from those of the convective
region below, the implications of which are discussed below. In the 2D
wedge with a hydro mixing initial model (ob.2d.m), the mixed region
stabilizes near the boundary predicted by the initial model. The lower
panel shows ob.2d.m with the boundaries of the initial model
indicated in yellow. The mixed region extends past the lower boundary
of the initial model mostly because of a resolution effect. Higher
resolution ameliorates the effect, and the qualitative behavior of the
boundaries is robust with changes in resolution \citep[Meakin \&
Arnett 2005, in prep.][]{ayr02}.

The hydrodynamic behavior observed in the simulations can be broken
down into three classes. The first regime is that of full
convection. Material is subject to engulfment by plumes and the flow
is highly turbulent. The convective boundary and radiative regions
comprise the second and thrid regimes, and can be roughly
characterized by the Richardson number, $Ri = N^2/(\delta u/\delta
r)^2$, where $N$ is the Br\"{u}nt-V\"{a}is\"{a}l\"{a} frequency, and
$\delta u/\delta r$ is the radial gradient of the shear
velocity. Broadly speaking, $Ri$ compares the kinetic energy in the
shear to the potential energy across a stratified layer. We will refer
to the region with $Ri \sim 0.25$ as the convective boundary
region. Material here is not
engulfed by rising plumes. Instead, plumes cause a Lagrangian
displacement of material, converting the kinetic energy of the
convective flow to internal wave energy. This energy conservation is
ignored in 1D treatments of convection, but turns out to have a
significant impact upon the structure of the star. The waves quickly
become non-linear and break. This region is also subject to shear
instabilities from plume spreading generating shear at the base of
this region. As a result, the boundary layer becomes well mixed, and
fresh fuel is entrained into the convective shell. Beyond this region
the waves are linear, and we enter the third regime. Dissipation of
the waves generates vorticity according to Kelvin's Theorem and
drives slow compositional mixing. These waves will also play a part in
generation of large angular scale perturbations in thermodynamic and
structural quantities, neutrino cooling, and intershell interactions,
though we consider only the impact of compositional mixing in this
paper.

The relatively close match between the 1D hydro mixing model and the
dynamic O shell simulations deserves further discussion. The area
inside the standard model boundary is fully convective, with a
velocity pattern characteristic of convective plumes. The additional
extent of mixing in ob.2d.c and ob.3d.M has a different velocity
pattern. The longitudinal banding in Figure~\ref{fig1} is
characteristic of wave motion. This is the low $Ri$ boundary region
where mixing is efficient. TYCHO evaluates a Richardson number for
shear in waves and plume spreading at the driving frequency of the
convection and mixes efficiently in regions with low Richardson
number. As long as the spatial extent of a low $Ri$ region produced by
driving at the overall convective turnover frequency is not
significantly smaller than that produced by an ensemble of plume
impacts carrying the same energy, the mixing predicted in 1D should be
similar to multi-D simulations. The caveat is that comparison of
simulations with with the hydro mixing model for O burning is limited
to a single time sequence from a single set of initial
conditions. More simulations are required to confirm that TYCHO
predicts the extent of the boundary region correctly.

Additional features are seen in multi-D which cannot be extended to
1D, but are interesting with regard to the progenitor. There is a
significant wave flux in the region between the O and C burning shells
with Mach numbers of several percent that gives rise to temperature
and density perturbations of order 0.1-1\%. Compressibility is an
important feature of the flow. Since the perturbations are generated
by internal waves, they have the potential to be correlated on large
angular scales. As we are dealing with wedges, we cannot comment on
how global these perturbations may be, save to say that the lowest
order wave modes allowed by the domain are present in
the simulations. Such large scale ``rippling'' within the progenitor
may provide the seeds for asymmetries in the explosion.

\section{CORE COLLAPSE MODELS AND EXPLOSIONS}

Three properties of the one-D models concern us most. The density and
entropy profiles of the core will determine the timing and energy of
the explosion \citep{fryer99}. In the neutrino-driven supernova model,
neutrinos heat material just beyond the proto-neutron star core. To
drive an explosion, this heated material must overcome the ram
pressure of the imploding star.  The success or failure of the
explosion mechanism is determined, in part, by the strength of this
ram-pressure which, in turn, depends upon the density in the region
between $\sim$1.5-2.0M$_\odot$.  The higher density for the progenitor
including hydro mixing means that it will require more energy to
explode. The explosion along with the star's abundance profile will
determine the final nucleosynthesis.

\begin{figure}
\figurenum{2}
\includegraphics[scale=0.8]{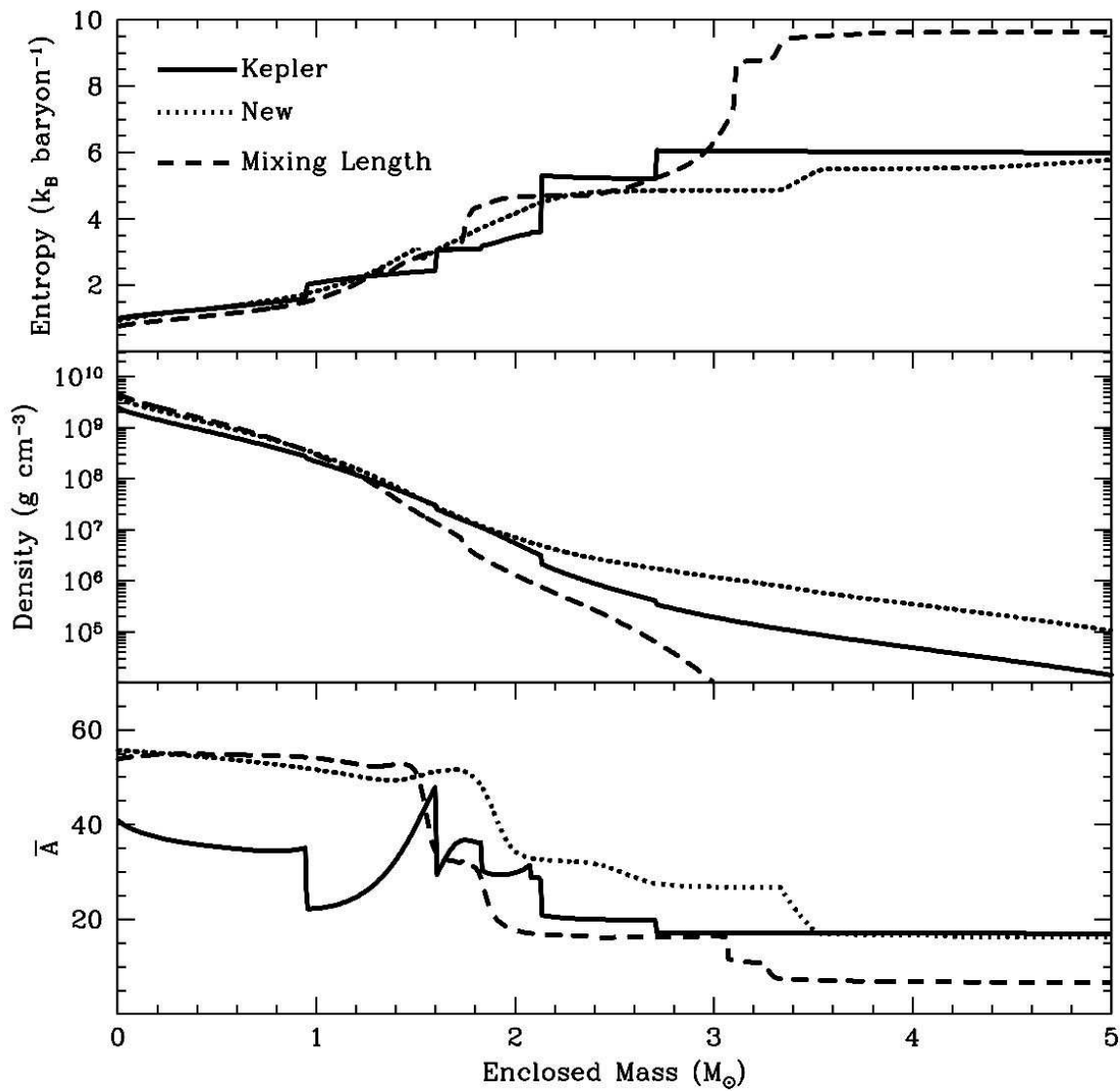}
\caption{\label{fig2} Entropy (top), density (middle), and mean atomic
weight (\={A}, bottom) versus enclosed mass for hydro mixing (dotted),
standard (dashed), and parametrized overshooting \citep{hw} (solid)
models. The primary differences in all three plots result from the
different \={A} $> 16$ core sizes. The entropy boundary at the outer
edge of the oxygen shell moves off the plot for the hydro mixing
model.}
\end{figure}

Figure~\ref{fig2} shows entropy (top), density (middle), and mean
atomic weight (\={A}, bottom) versus enclosed mass for the TYCHO core
collapse models and a 23 \sol Kepler model \citep{hw} with
parametrized overshooting calibrated for lower mass stars
\citep{wzw78}. Several differences are apparent. Sharp entropy
gradients are present in the mixing length and Kepler models. Entropy
and species are transported by dissipative waves, and these gradients
are somewhat smoothed out. Second, the density in the hydro mixing
model is consistently higher.  The density in the Kepler model with
parametrized overshooting is an order of magnitude lower that the wave
physics model outside of 2.5 \sol, and the situation is even more
extreme for the standard model. The final mass dominated by species of
$A\ge 16$ is twice as large as the mixing length model. Finally, the
extent of the core which has been processed by different burning
stages (indicated by \={A}) is larger.

We calculate three explosions using an updated version of the core
collapse code described in \citet{fryer99}.  We have artificially scaled
up the luminosity of the neutrinos beyond the neutrinosphere to induce
an explosion.  We do this by setting the free-streaming boundary for the
light-bulb approximation of the \citet{fryer99} code to the neutrinosphere
and artificially raising the luminosity of both the electron and anti-electron
neutrinos at this boundary.  This scaling is roughly 20\% above a failed
1-dimensional model for our standard model and the weak wave physics
model.  In the strong explosion wave-physics model, we increase the
neutrino luminosity by another 40\%.  The raised neutrino luminosities
alter the structure of the collapsed core, causing the observed luminosities
to not vary as much as our scaling factor, but the explosion energetics
change dramatically.  The energetics, remnant masses, and several
interesting abundance ratios are given in Table~\ref{tab2}. The ratio
$\alpha /$Fe is defined in terms of the network as
\begin{equation}
\frac{\sum {\rm O,Ne,Mg,Si,S,Ar,Ca,Ti,Cr}}{\sum {\rm Fe,Ni}}
\end{equation}


The difference between the models is striking. For the same neutrino
luminosity, the explosion energy of the hydro model drops by 65\%, and
the remnant mass is much larger. It requires an exceptionally powerful
explosion to produce close to the same remnant mass. This energetic
event would produce a very large mass of Ni. Interstingly, $\alpha /$
Fe is similar for the standard model and the energetic explosion of
the hydro mixing model. The main difference aside from the total mass
of ejected material is the C/O ratio, which is much lower in the hydro
mixing model. The $\alpha /$ Fe for the weak explosion should be read
merely as ``large'' since mixing during the explosion may increase the
amount of Fe ejected to a few hundredths of a solar mass.

Though the mixing in late burning stages is currently unconstrained by
direct observations, we may draw some general conclusions. Even if no
mixing beyond the standard model were to take place after the main
sequence, the apsidal motion test indicates that the core will be
larger than the standard model. Overshooting models do produce larger
cores, but since most schemes are empirically calibrated on lower mass
stars, they are not predictive, and diverge increasingly from real
stars with increasing stellar mass in the sense of underestimating
core sizes. Both the hydrodynamic simulations and basic physical
consistency argue for the mixing processes to continue in late burning
stages, enlarging the core beyond the apsidal motion limit. We use a
predictive theory for this hydrodynamic mixing in our evolution code,
which is informed by multi-D simulations. Conservatively, the
difference in core sizes between standard, calibrated overshooting,
and hydro mixing models demonstrates the uncertainty in supernova
calculations arising from progenitor models.  

Though the changes in the progenitors are large, it is difficult to
predict what the integrated effect on a population will be. Larger
core sizes result in higher stellar luminosities, and the path of very
high mass stars through the HR diagram is qualitatively very
different. The IMF derived from combining models with observed
luminosity functions will change. The frequency of different mass
supernova progenitors and their contribution to nucleosynthetic yields
will change accordingly. A full synthetic population will be required
to determine the impact of the change in progenitor models upon
integrated yields.

\acknowledgements This work was funded in part under the auspices of
the U.S.\ Dept.\ of Energy, and supported by its contract
W-7405-ENG-36 to Los Alamos National Laboratory, by a DOE SciDAC grant
DE-FC02-01ER41176, an NNSA ASC grant, and a subcontract to the ASCI
FLASH Center at the University of Chicago.

\clearpage

\begin{deluxetable}{crrrrrrr}
\tabletypesize{\scriptsize}
\tablecaption{Oxygen Shell Burning Models \label{tab1}}
\tablewidth{0pt}
\tablehead{
\colhead{Model} & \colhead{Hydro} & \colhead{$r_{in}$} & \colhead{$r_{out}$}  & \colhead{$\Delta \phi$} & \colhead{$\Delta \theta$} & \colhead{$N_r \times N_{\phi} \times N_{\theta}$} & \colhead{$t_{max}$} \\
\colhead{} & \colhead{Mixing} & \colhead{cm} & \colhead{cm} & \colhead{} & \colhead{} & \colhead{} & \colhead{sec}}
\startdata
ob.3d.B & n & 0.3 & 1.0 & 30 & 30 & $400 \times 100 \times 100$ & 64 \\
ob.2d.c & n & 0.3 & 1.0 & 90 & \nodata & $400 \times 320$ & 574 \\
ob.2d.m & y & 0.3 & 1.0 & 90 & \nodata & $400 \times 320$ & 800 \\
\enddata
\end{deluxetable}

\clearpage

\begin{deluxetable}{crrrrrrrr}
\tabletypesize{\scriptsize}
\tablecaption{Explosion Calculations \label{tab2}}
\tablewidth{0pt}
\tablehead{
\colhead{Model} & \colhead{$E_{expl}$} & \colhead{$M_{rem}$}  & \colhead{$M_{Ni}$} & \colhead{C/O} & \colhead{O/Fe} & \colhead{Si/Fe}  & \colhead{Ti/Fe} & \colhead{$\alpha /$Fe} \\
\colhead{} & \colhead{f.o.e.} & \colhead{\sol} &\colhead{\sol} & \colhead{} & \colhead{} & \colhead{} & \colhead{} & \colhead{}}
\startdata
standard & 1.65 & 1.57 & 0.42 & 0.94 & 4.05 & 0.48 & $2.1^{-3}$ & 5.74\\
hydro mixing & 0.57 & 6.01 & $4.0^{-4}$ & 0.19 & 463 & 0.60 & $2.9^{-3}$ & 528\\
hydro mixing & 3.0 & 1.64 & 0.99 & 0.12 & 4.53 & 0.62 & $1.4^{-3}$ & 6.05\\
\enddata
\end{deluxetable}
\clearpage

\clearpage

\clearpage

\begin{thebibliography}

\bibitem[Alexakis et al.(2002)]{ayr02} Alexakis, A., Young, Y., \&
Rosner, R. 2002, \prd, 65, 026313

\bibitem[Arnett(1996)]{arnett96} Arnett, David 1996, {\it Supernovae
and Nucleosynthesis}, Princeton University Press

\bibitem[Asida \& Arnett(2000)]{sa00}Asida, S.M., \& Arnett, D. 2000 
   \apj, 545, 435

\bibitem[Asplund et al.(2005)]{as05} Asplund, Martin, Grevesse, N. \& Sauval,
A. J. 2005, astro-ph 0410214

\bibitem[Baz\`{a}n \& Arnett(1998)]{ba98} Baz\`{a}n, G., \& Arnett, D. 
1998, \apj, 494, 316

\bibitem[Grevesse \& Sauval(1998)]{gs98} Grevesse, N. \& Sauval,
A. J., 1998, Space Science Reviews, 85, 161

\bibitem[Fitts, Vadas, \& \O yvind(1998)]{fva98} Fitts, David C.,
Vadas, Sharon L., \& Andreassen, \O yvind 1998, \aap, 333, 343

\bibitem[Fryer(1999)]{fryer99} Fryer, Christofer L. 1999, \apj, 522, 413

\bibitem[Nordlund \& Stein(2001)]{ns01} Nordlund, \AA. \& Stein,
R. F. 2001, \apj, 546, 576

\bibitem[Press(1981)]{press81} Press, W. H. 1981, \apj, 245, 286

\bibitem[Rauscher et al.(2002)]{hw} Rauscher, T., Heger, A., Hoffman,
R. D., \& Woosley, S. E. 2002, \apj, 576, 323

\bibitem[Stein \& Nordlund(2001)]{sn01} Stein, R. F. \& Nordlund,
\AA. 2001, \apj, 546, 585

\bibitem[Weaver, Zimmerman \& Woosley(1978)]{wzw78} Weaver, T. A.,
Zimmerman, G. B., \& Woosley, S. E. 1978, \apj, 225, 1021

\bibitem[Young et al.(2005)]{ya05} Young, Patrick A. \& Arnett, David
2005, \apj, 618, 908

\bibitem[Young et al.(2003)]{ykra03} Young, Patrick A., Knierman, Karen A.,
Rigby, Jane R., \& Arnett, David 2003, \apj, 595, 1114

\bibitem[Young, Mamajek, Arnett, \& Liebert(2001)]{ymal01} Young, P. A., 
Mamajek, E.~E., Arnett, D., \& Liebert, J. 2001, \apj, 556, 230

\end{thebibliography}
\end{document}